\documentclass[namedreferences,hyperref,optionalrh,solaromanenum]{spr-sola}
\usepackage{graphicx}                    
\usepackage{color}                       

\chardef\us=`\_

\newcommand{\arcsec}{$^{\prime\prime}$}
\usepackage{color}

\begin{document}

\begin{frontmatter}



\title{Comparison of Polar Magnetic Fields Derived from MILOS and MERLIN Inversions with Hinode/SOT-SP Data}

%


\author[addressref={aff1,aff2},email={masahito.kubo@nao.ac.jp}]{\inits{M.}\fnm{Masahito}~\lnm{Kubo}\orcid{0001-5616-2808}}
\author[addressref=aff3,email={}]{\inits{D.}\fnm{Daikou}~\lnm{Shiota}\orcid{0002-9032-8792}}
\author[addressref={aff1,aff2},email={}]{\inits{Y.}\fnm{Yukio}~\lnm{Katsukawa}\orcid{0002-5054-8782}}
\author[addressref={aff1,aff2},email={}]{\inits{M.}\fnm{Masumi}~\lnm{Shimojo}\orcid{0002-2350-3749}}
\author[addressref={aff4,aff5},email={}]{\inits{D.}\fnm{David}~\lnm{Orozco Su{\'a}rez}\orcid{0001-8829-1938}}
\author[addressref=aff6,email={}]{\inits{N.V.}\fnm{Nariaki}~\lnm{Nitta}\orcid{0001-6119-0221}}
\author[addressref=aff6,email={}]{\inits{M.}\fnm{Marc}~\lnm{DeRosa}\orcid{0002-6338-0691}}
\author[addressref=aff7,email={}]{\inits{R.}\fnm{Rebecca}~\lnm{Centeno}\orcid{0002-1327-1278}}
\author[addressref=aff8,email={}]{\inits{H.}\fnm{Haruhisa}~\lnm{Iijima}\orcid{0002-1007-181X}}
\author[addressref=aff8,email={}]{\inits{T.}\fnm{Takuma}~\lnm{Matsumoto}\orcid{0002-1043-9944}}
\author[addressref=aff8,email={}]{\inits{S.}\fnm{Satoshi}~\lnm{Masuda}\orcid{0001-5037-9758}}

%

\runningauthor{Kubo et al.}
\runningtitle{Comparison of polar magnetic fields}


\address[id=aff1]{National Institutes of Natural Sciences, National Astronomical Observatory of Japan, 2-21-1 Osawa, Mitaka, Tokyo, 181-8588, Japan}
\address[id=aff2]{The Graduate University for Advanced Studies (SOKENDAI), 2-21-1 Osawa, Mitaka, Tokyo, 181-8588, Japan}
\address[id=aff3]{National Institute of Information and Communications Technology, Koganei, Tokyo, 184-8795, Japan}
\address[id=aff4]{Instituto de Astrof\'{i}sica de Andaluc\'{i}a, Glorieta de la Astronom\'{i}a s/n, 18008, Granada, Spain}
\address[id=aff5]{Spanish Space Solar Physics Consortium (S$^3$PC)}
\address[id=aff6]{Lockheed Martin Solar and Astrophysics Laboratory, Palo Alto, CA 94304, USA}
\address[id=aff7]{National Center for Atmospheric Research, High Altitude Observatory, Boulder, CO, 80307, USA}
\address[id=aff8]{Institute for Space-Earth Environmental Research, Nagoya University, Nagoya, Aichi, 464-8601, Japan}

\begin{abstract}
The detailed investigation of the polar magnetic field and its time evolution is one of the major achievements of Hinode. 
Precise measurements of the polar magnetic field are essential for understanding the solar cycle, as they provide important constraints for identifying the source regions of the solar wind. 
The Spectropolarimeter (SP) of the Solar Optical Telescope (SOT) on board Hinode has been the instrument best suited to make such measurements. 
In this study, we compare the SOT-SP data for the polar regions, processed using two representative Milne-Eddington inversion codes, MILOS and MERLIN. 
These codes are applied to the same level-1 SOT-SP data, and the same disambiguation algorithm is used on the maps that go through the two inversions. 
We find that the radial magnetic-flux density (the magnetic-flux density with respect to the local vertical) provided by the MERLIN inversion tends to be approximately 7\%-10\% larger than that obtained from the MILOS inversion. 
The slightly higher radial magnetic-flux density from MERLIN appears to be common to the polar magnetic fields observed at different phases of the solar cycle. 
When MILOS is run with the same scattered-light profile and the same magnetic filling factor that are derived with the MERLIN inversion, the radial magnetic-flux density derived from the two inversions is almost the same. 
We attribute the difference in the radial magnetic-flux density to different filling factors adopted by the two inversions, based on whether the scattered-light profiles are assumed to be the Stokes I profiles averaged over the neighboring pixels or over the entire field of view. 
The relationship between the radial magnetic-flux density and magnetic filling factor could be more complex in the polar (limb) observations due to the possible contributions of the transverse magnetic-field component to the estimation of the radial magnetic-flux density.
\end{abstract}

%


\end{frontmatter}

%

\section{Introduction}
     \label{S-Intro} 

Observations of the magnetic field in the solar polar regions are important for understanding the solar cycle variations. 
In particular, the magnetic flux in the polar region is considered a good indicator for predicting the level of activity of the next cycle because it is well correlated with the maximum sunspot number in the next cycle \citep{2010LRSP....7....3C, 2010LRSP....7....1H, 2015LRSP...12....5P}. 
Initially, because the polar magnetic field was difficult to measure, the spatial distribution of the magnetic flux in the polar regions was inferred from the observations of the polar faculae \citep[e.g.][]{1964ApJ...140..731S, 1991ApJ...374..386S}. 
Since the 1960s, the magnetic flux in the polar regions has been directly monitored using full disk line-of-sight magnetograms from various observatories (e.g., Wilcox Solar Observatory, Mt. Wilson Observatory, Kit peak Observatory, SoHO/MDI).
However, it is challenging to accurately measure the magnetic flux in the polar region above 70$^\circ$ latitude (which is likely concentrated as discrete elements as suggested by the polar faculae) using the full disk line-of-sight magnetograms because of the limited spatial resolution near the limb, which limits access to the weakest magnetic signals. In addition, the limb darkening effect reduces the signal-to-noise ratio, thereby limiting the polarimetric sensitivity. 
Another reason is the specific distribution of the polar fields which makes linear polarization measurements more relevant.

High spatial resolution and high precision polarization data can be obtained from the solar polar region using the Spectropolarimeter \citep[SP:][]{2013SoPh..283..579L} of the Solar Optical Telescope \citep[SOT:][]{2008SoPh..249..167T} onboard the Hinode satellite launched in September 2006 \citep{2007SoPh..243....3K}.
These polarization data have enabled, for the first time, the detailed investigations of the photospheric vector magnetic field in the polar region. 
Prior to the SOT-SP observations, magnetic-field measurements of the polar faculae with the strong magnetic fields have been carried out on the ground with the highly sensitive spectropolarimeter \citep{1997SoPh..175...81H, 2004A&A...425..321O, 2007A&A...474..251B}. 
In these studies, only the line-of-sight magnetic field is inferred from the circular polarization measurements.
The SOT-SP observations show the properties of the vector magnetic field for the KG patches widely scattered in the polar regions \citep{2008ApJ...688.1374T}.
The ubiquitous horizontal magnetic fields discovered in the quiet region near the disk center \citep{2007ApJ...666L.137C, 2008ApJ...672.1237L, 2008A&A...481L..25I} can also be detected in the high-latitude region \citep{2010ApJ...719..131I}. 
The detailed process of magnetic-field reversal in the polar region was obtained from the regular polar observations by Hinode/SOT-SP over the period of one solar cycle \citep{2012ApJ...753..157S, 2022ApJ...941..142P}.
The observations from the monitoring of the polar regions revealed that polar field reversal is caused by magnetic elements with large magnetic-flux density, such as kilogauss patches, and that magnetic elements with small magnetic flux, such as the ubiquitous horizontal magnetic fields, show little variation in the polar field \citep{2012ApJ...753..157S}. 

Our understanding of the polar magnetic field has thus improved significantly owing to the SOT-SP observations.
However, we need to know how accurate the SOT-SP measurements are for the magnetic flux in the polar regions, which may have implications, for example, in the so-called “open flux problem” \citep{2017ApJ...848...70L}. 
It was shown that the open magnetic flux calculated from the photospheric magnetograms is much lower than the observed in situ at 1 AU from the Sun as far as data during near solar minimum are concerned.
It is important to find out how much of these discrepancies may be accounted for by the possible inadequacy of measuring the magnetic flux in the polar regions. In order to address the uncertainties in the SOT-SP data analysis with the Milne-Eddington model atmospheres for the polar magnetic flux, this article compares magnetograms of the polar regions that are processed using two representative Milne-Eddington inversion codes, MILne-Eddington inversion of pOlarized spectra \citep[MILOS:][]{2007A&A...462.1137O} and Milne-Eddington gRid Linear Inversion Network \citep[MERLIN:][]{2007MmSAI..78..148L}. While MERLIN is used to produce the official level-2 database at the Community Spectro-polarimetric Analysis Center (CSAC) of the High Altitude Observatory (HAO)\footnote{\url{csac.hao.ucar.edu/sp_data.php}}, MILOS has been often used for the studies of the polar magnetic field \citep{2008ApJ...688.1374T, 2010ApJ...719..131I, 2012ApJ...753..157S, 2013ApJ...776..122K, 2015ApJ...799..139K}.
Furthermore, MILOS is also used for the new SOT-SP database of the polar regions at the Hinode Science Center of the Institute for Space-Earth Environmental Research (ISEE), Nagoya University\footnote{\url{hinode.isee.nagoya-u.ac.jp/sot_polar_field/}, doi: 10.34515/DATA.HSC-00001}. 
In Section~\ref{S-data}, we briefly describe the procedure of our analysis and the results are presented in Section~\ref{S-results}.  We discuss and summarize our findings in Section~\ref{S-Summary}.

\section{Data and Analysis}
     \label{S-data}
\subsection{Polar Magnetic Field Observed by Hinode (HOP 206)}      

\begin{figure} 
\centerline{\includegraphics[width=12cm]{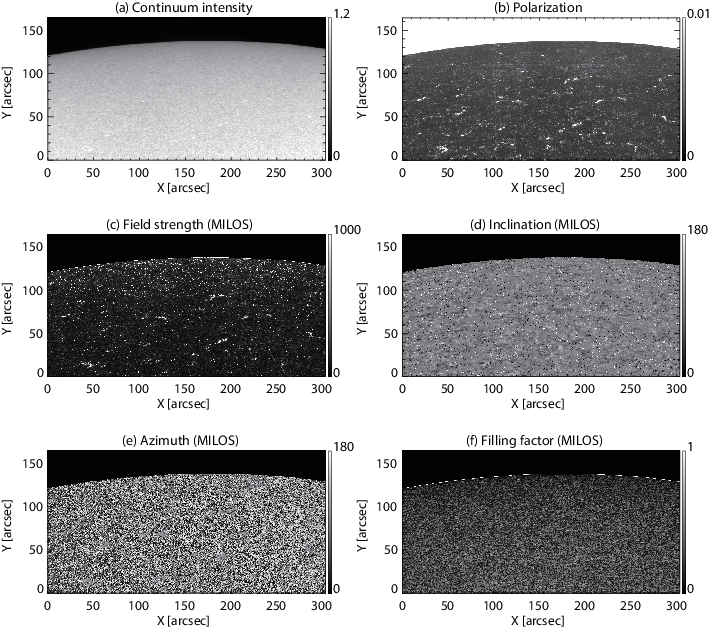}}
\caption{Example of the polar observation with Hinode/SOT-SP for (a) continuum intensity, (b) degree of polarization, (c) magnetic-field strength, (d) magnetic-field inclination, (e) magnetic-field azimuth, and (f) magnetic filling factor.
The magnetic-field parameters (c-f) are derived using MILOS inversion.
The magnetic-field inclination is defined in the line-of-sight frame: 0 degree when the magnetic field is along the line-of-sight and directed toward the observer, and 180 degree when the line-of-sight magnetic field is directed away from the observer.
The observation was obtained from the north polar region on 23 August 2021.
The \textit{horizontal} and \textit{vertical axes} are shown in units of arcseconds.
}
\label{F-vector_map}
\end{figure}

\begin{figure} 
\centerline{\includegraphics[width=10cm,clip=]{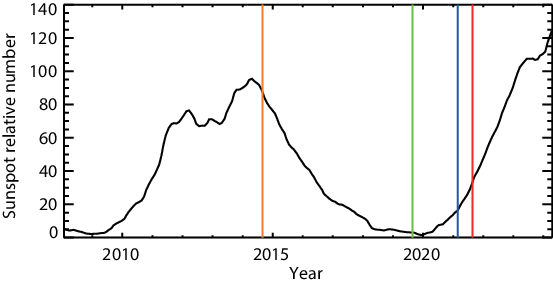}}
\caption{Thirteen-month running mean of sunspot relative number obtained by the Solar Science Observatory at the National Astronomical Observatory of Japan.  
The \textit{vertical lines} correspond to the dates in the dataset shown in Table~\ref{T-dataset}.}
\label{F-solar_cycle}
\end{figure}

\begin{table}
\caption{Hinode dataset (HOP 206).}
\label{T-dataset}  
\begin{tabular}{ccccc}     
\hline                   
Date & Start time & End time & Location & Solar cycle phase\\
\hline
28-Aug-2014 & 13:28:07UT & 16:19:22UT  & North pole & Maximum\\
25-Aug-2019 & 12:59:03UT & 15:52:32UT  & North pole & Minimum\\
26-Feb-2021 & 11:08:08UT & 14:01:36UT  & South pole & Intermediate\\
23-Aug-2021 & 10:45:05UT & 13:38:34UT  & North pole & Intermediate\\
\hline
\end{tabular}
\end{table}

The Hinode team carries out campaign observations of the north polar region at 3-day intervals for one month during August-September and the south polar region for one month during February-March.
This campaign, called HOP 206, has been conducted regularly since 2012 to obtain polar panoramic maps of the magnetic field.
Figure~\ref{F-vector_map} shows an example of the HOP 206 observation.
During this observation period, SOT-SP scanned the polar region with a field of view of 300\arcsec$\times$164\arcsec.
The pixel scale in the scan direction was 0.30\arcsec and that along the slit was 0.32\arcsec, corresponding to 2$\times$2 binnings of the original SOT-SP sampling.
The integration time was 9.6 s at each scan position and the total time to acquire one map was approximately 2.7 h.
The full polarization states of two photospheric Fe {\footnotesize I} lines at 630.15 nm and 630.25 nm along the slit were measured at each scan position.

In this study, four datasets of the HOP 206 are examined, as shown in Table~\ref{T-dataset}. 
The first and second datasets are obtained for the north polar regions in the solar maximum and solar minimum periods, respectively (Figure~\ref{F-solar_cycle}).
The third and fourth datasets are  observed in the south and north polar regions, respectively, in the intermediate phase between the solar minimum and maximum.


\subsection{Comparison of the MILOS and MERLIN Inversions}
In this study, the vector field maps of the polar regions derived by the MILOS inversion are compared with those derived by the MERLIN inversion.
The MERLIN inversion results are archived in the SOT-SP level-2 database at HAO.
We produce MILOS inversions from the level-1 data, which are Stokes profiles calibrated with the standard procedures \citep{2013SoPh..283..601L}  of the HAO database.
The 180$^\circ$ ambiguity of the azimuth angle is resolved using the method in \cite{2010ApJ...719..131I} in both the MILOS and MERLIN inversions.
The differences in the comparative results of this study therefore can be attributed to the differences between the MILOS and MERLIN inversions.
The disambiguation method in \cite{2010ApJ...719..131I} defines three groups: vertical (0$^\circ$-40$^\circ$ or 140$^\circ$-180$^\circ$), horizontal (70$^\circ$-110$^\circ$), undetermined (40$^\circ$-70$^\circ$ or 110$^\circ$-140$^\circ$ ) fields, depending on the two magnetic-field configurations to which the ambiguity leads to in the local reference frame. 
In this study, the pixels are excluded from the comparison if one of the results with the inversion codes selects the undetermined field  group in the final configuration. 
All the pixels defined in the vertical/horizontal field groups are compared even if the pixel is defined in the opposite group. 
The comparison between these two inversions was done in \cite{2023ApJ...951...23C} by using a numerical simulation of a plage region at the inclined viewing angle of 65$^\circ$.

The MILOS and MERLIN inversions use a least-squares inversion technique based on Milne-Eddington atmospheres.
The MILOS inversion has 10 free parameters: magnetic-field strength, inclination angle, azimuth angle, Doppler shift, Doppler width, damping parameter, ratio of line center to continuum opacity, slope and surface value of the source function, and magnetic filling factor.
The MERLIN inversion treats the Doppler shifts (line centers) of the two Fe I lines as two independent free parameters instead of a common Doppler shift (one free parameter) of the two lines.
The other nine free parameters used in the MILOS inversion are also used in the MERLIN inversion.
Furthermore, the Doppler shift of a scattered-light profile is an additional free parameter in the MERLIN inversion.
The scattered-light profile is used to estimate the magnetic filling  factor.
In both inversions, the observed Stokes profiles are fitted with 
\begin{equation}
\label{E-ff}
\left(
\begin{array}{c}
I \\
Q \\
U \\
V
\end{array}
\right)
=f
\left(
\begin{array}{c}
I_{\rm{mag}} \\
Q_{\rm{mag}} \\
U_{\rm{mag}} \\
V_{\rm{mag}}
\end{array}
\right)
+(1-f)
\left(
\begin{array}{c}
I_{\rm{scatt}} \\
0 \\
0 \\
0
\end{array}
\right),
\end{equation}
where $(I_{\rm{mag}}, Q_{\rm{mag}}, U_{\rm{mag}}, V_{\rm{mag}})$ are the magnetized components, $I_{\rm{scatt}}$ is the scattered-light profile, and $f$ is the magnetic filling factor. 
The assumption of the scattered-light profile is different between the MILOS and MERLIN inversions (see Section~\ref{S-ff}).

\section{Results}
     \label{S-results}

\subsection{Radial Magnetic Flux Density}           
\label{S-flux}
\begin{figure} 
\centerline{\includegraphics[width=12cm]{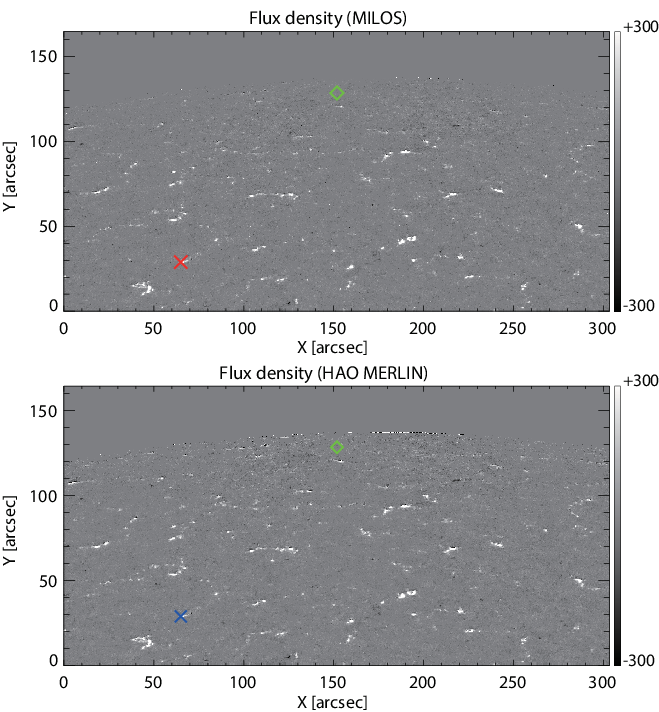}}
\caption{Example map of the radial magnetic-flux density derived with the MILOS (\textit{top}) and  MERLIN (\textit{bottom}) inversions on 23 August 2021.
The symbols indicate the positions for which the scattered-light  profiles are shown in Figure~\ref{F-stray}.
} 
\label{F-flux_map}
\end{figure}


\begin{figure} 
\centerline{\includegraphics[width=12cm]{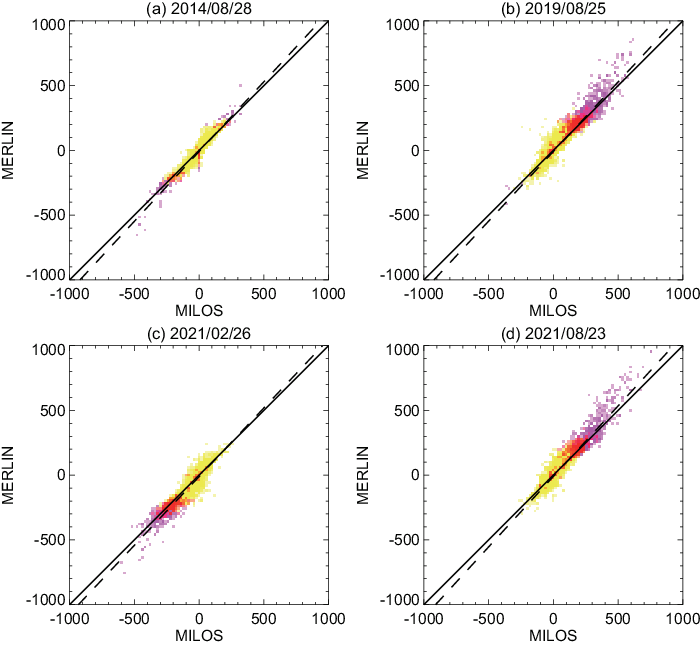}}
\caption{Two-dimensional histograms of the radial magnetic-flux density derived with the MILOS vs. MERLIN inversions for the four datasets shown in Table~\ref{T-dataset}. 
The magnetic-flux density values are shown in units of Gauss.
The distributions of the pixels with degree of polarization greater than 0.01 and between 0.01 and 0.004 are displayed in \textit{magenta} and \textit{yellow colors}, respectively. 
The overlap of the \textit{magenta} and \textit{yellow color} distributions is shown in \textit{red}.
The \textit{solid line} denotes equal inversion results.
The \textit{dashed line} indicates the result of the linear fit, and its slope is 1.09, 1.09. 1.07, and 1.10 from panels (a) to (d).
}
\label{F-flux}
\end{figure}

Figure~\ref{F-flux_map} shows the maps of the radial magnetic-flux density, which is the magnetic-flux density with respect to the local vertical, in the north polar region on 23 August 2021. 
This value is estimated using the method proposed by \cite{2012ApJ...753..157S}. 
The radial magnetic-flux density is defined as $fB_{r}$, where $f$ is the magnetic filling factor and $B_r$ is the radial component of the magnetic-field vector.
The flux density maps obtained by the MILOS and MERLIN inversions appear similar. 
Strong magnetic patches \citep{2008ApJ...688.1374T} can be seen and positive magnetic polarity is dominant in the north polar region.
Figure~\ref{F-flux} shows the scatter plot of the radial magnetic-flux density obtained with the two inversions on the datasets.
The scatter plots show that the radial magnetic-flux density tends to be approximately 1.1 times larger in the MERLIN inversion results than in the MILOS inversion results.
This is more evident for the pixels with strong polarization (magenta symbols in Figure~\ref{F-flux}).
The degree of polarization is estimated by $\sqrt(Q^2+U^2+V^2)/I_c$ integrated over the two Fe I lines.
The trend of about 7-10 \% higher magnetic-flux density in the MERLIN inversion results is observed in the datasets obtained at different phases of the solar cycle, as shown in Figure~\ref{F-flux}, although the number of strong magnetic patches varies with the solar cycle.
It is also confirmed that the results are almost the same in the north and south polar regions, where the polarities of the magnetic-flux density are opposite. 
The common results for the different datasets suggest that the difference in the radial magnetic-flux density is a characteristic of each inversion.
In the following sections, we investigate the origin of this difference in the magnetic-flux density obtained with the MILOS and MERLIN inversions by using the dataset on 23 August 2021.

\subsection{Dependence of Scattered-light Profile}
\label{S-ff}

\begin{figure} 
\centerline{\includegraphics[width=6cm]{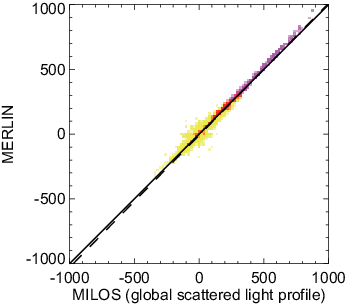}}
\caption{Two-dimensional histograms of MILOS vs. MERLIN for radial magnetic-flux density when the MILOS inversion is performed with the ``global'' scattered-light profile and the fixed filling factor equal to those used in the MERLIN inversion for each pixel. 
The distributions of the pixels with degree of polarization greater than 0.01 and between 0.01 and 0.004 are displayed in \textit{magenta} and \textit{yellow colors}, respectively. The overlap of the \textit{magenta} and \textit{yellow color} distributions is shown in \textit{red}.
The \textit{solid line} in (b) denotes the instances where the two inversion results are equal whereas the \textit{dashed line} denotes the result of the linear fit. The slope of the \textit{dashed line} is 1.03.
}
\label{F-fixed_ff}
\end{figure}

\begin{figure} 
\centerline{\includegraphics[width=8cm]{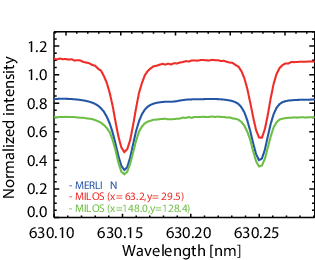}}
\caption{Scattered-light profiles used to estimate the magnetic filling factor for the MERLIN (\textit{blue}) and MILOS (\textit{red} and \textit{green}) inversion codes.
The \textit{blue profile} was downloaded from the HAO database.
The \textit{red} (\textit{green}) \textit{line} represents the profile at the position indicated by the \textit{red cross} (\textit{green diamond}) \textit{symbol} in Figure~\ref{F-flux_map}.
These scattered-light profiles are normalized by the reference continuum intensity, which is averaged over a specific region defined as in  \cite{2012ApJ...753..157S}. 
}
\label{F-stray}
\end{figure}

One major difference between the MERLIN and MILOS inversions is the assumption of the scattered-light profiles. The MERLIN inversion assumes the ``global'' scattered-light profile, which is averaged over the Stokes I profiles in the areas where the degree of polarization is weaker than 0.35\% in the entire field-of-view within the disk. The scattered-light profile is common to all input pixels of the inversion, but the Doppler shift of the scattered-light profile is one of the free parameters in the MERLIN inversion. In contrast, the MILOS inversion assumes a ``local'' scattered-light profile, which is averaged over the Stokes I profiles in the neighboring 7 $\times$ 7 pixels. The averaging process is not limited to the weak polarization pixels. The size of 7 pixels corresponds to 2.2{\arcsec} in the HOP 206 datasets. The scattered-light profile is different for each pixel and its Doppler shift is not a free parameter in the MILOS inversion.

We investigate how this difference affects the estimation of the radial magnetic-flux density. The second term in Equation~\ref{E-ff} consists of the scattered-light profile and the magnetic filling factor. We run the MILOS inversion with the ``global'' scattered-light profile and the magnetic filling factor fixed at the value derived with the MERLIN inversion. The ``global'' scattered-light profile available in the HAO database is shifted in the wavelength direction by using the fitting results of the MERLIN inversion for each pixel. As a result, the second term in Equation~\ref{E-ff} is the same as in the MERLIN inversion. In this case, the discrepancy of the radial magnetic-flux densities derived from the MILOS and MERLIN inversions is significantly reduced, especially in the areas of strong polarization (Figure~\ref{F-fixed_ff}). This result implies that the discrepancy in the magnetic-flux density is caused by the different assumption of the scattered-light profile.

Figure~\ref{F-stray} shows the scattered-light profile in the MERLIN inversion and two examples related to the MILOS inversion. 
These profiles are normalized by the same value.
The red profile is obtained at the strong magnetic patch (red symbol in Figure~\ref{F-flux_map}) and the green profile is obtained at the pixel very close to the solar limb (green symbol in Figure~\ref{F-flux_map}).
Their intensities are different due to the limb, and the gap between the continuum intensities is clearly seen.
The intensity level of the scattered-light profile used in the MERLIN inversion is between those of the red and green profiles.

\subsection{Magnetic Field Vector}
\label{S-vector}

\begin{figure} 
\centerline{\includegraphics[width=12cm]{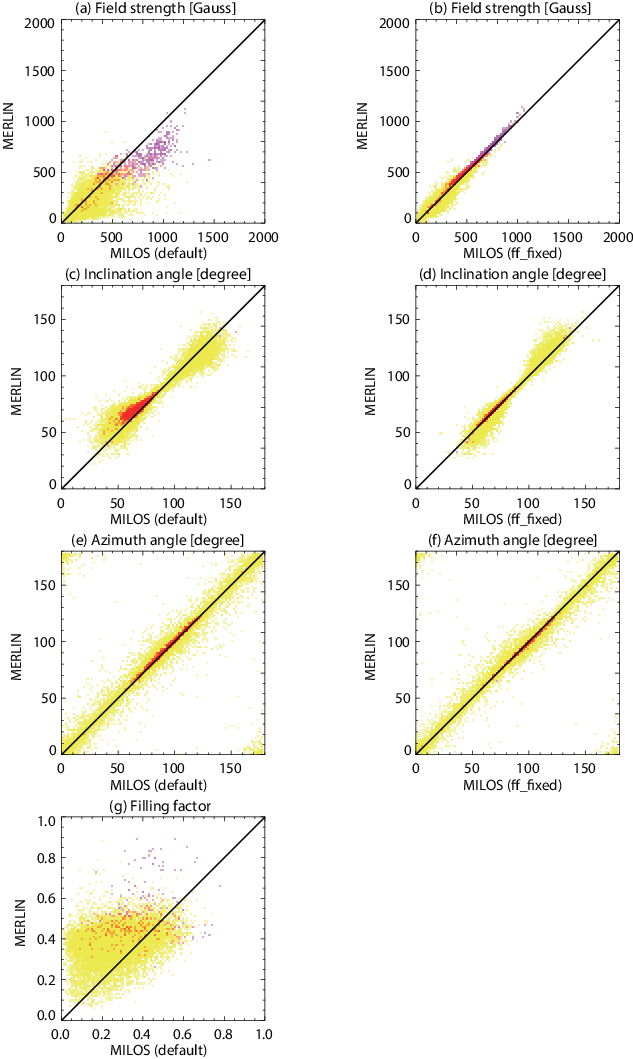}}
\caption{Two-dimensional histograms of MILOS vs. MERLIN for (a)/(b) magnetic-field strength, (c)/(d) magnetic-field inclination, (e)/(f) magnetic-field azimuth, and (g) magnetic filling factor. 
The magnetic-field inclination is defined in the line-of-sight frame as defined in the caption of Figure~\ref{F-vector_map}.
The \textit{left panel} represent the fitting results of the MILOS inversion with the default settings. 
The \textit{right panel} represent the results of the MLOS inversion with the ``global'' scattered-light profile and with the fixed filling factor equal to the MERLIN value.
The \textit{solid line} denotes the instances where the two inversion results are equal.
The distributions of the pixels with degree of polarization greater than 0.01 and between 0.01 and 0.004 are displayed in \textit{magenta} and \textit{yellow colors}, respectively. 
The overlap of the \textit{magenta} and \textit{yellow color} distributions is shown in \textit{red}.
}
\label{F-vector}
\end{figure}

A comparison of the magnetic-field vectors in the MILOS and MERLIN inversion results is shown in Figure~\ref{F-vector}.
In the areas of strong polarization (magenta symbols), the results of the two inversions have a nearly linear relationship for the magnetic-field strength, inclination angle, and azimuth angle. 
The difference of approximately 180$^\circ$ in the plot for the azimuth angle is attributed to azimuth ambiguity.
The deviation from the linear relationship becomes smaller for the magnetic-field strength and inclination angle when the same ``global'' scattered-light profile and the same filling factor are used in these two inversions (right panels in Figure~\ref{F-vector}). The inversion with the ``global'' scattered-light profile tends to provide slightly weaker field strength, larger magnetic filling factor, and more inclined magnetic fields. 
 The comparison of the magnetic-field vector between the two inversions with their default settings agrees with the results of the numerical simulations \citep{2023ApJ...951...23C}.

\subsection{Fitting Results}
\begin{figure} 
\centerline{\includegraphics[width=12cm]{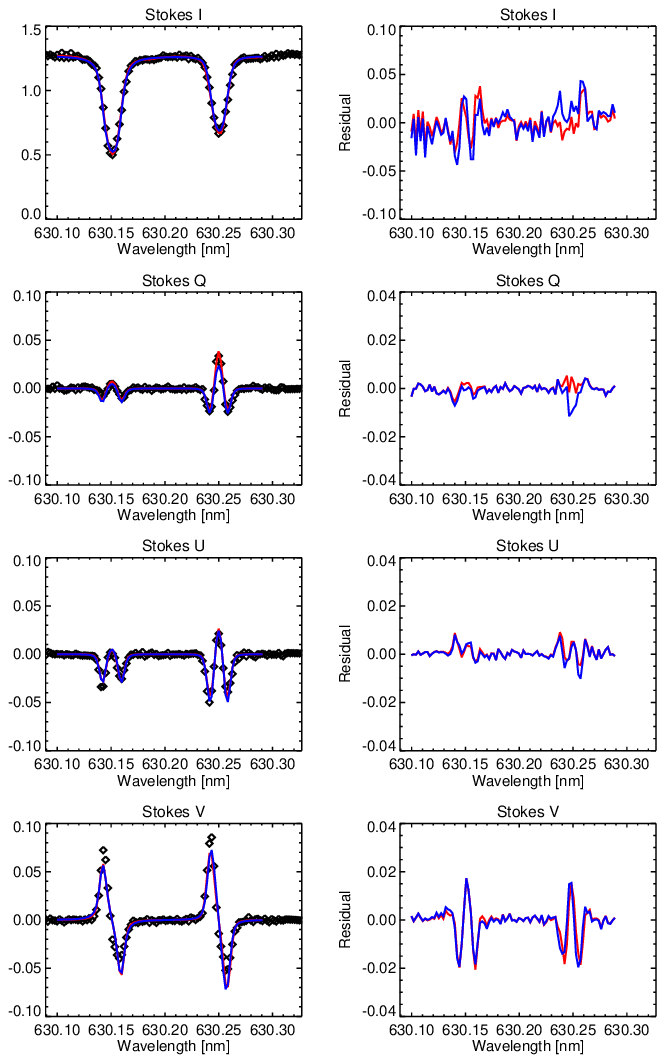}}
\caption{
Examples of the Stokes profiles (\textit{left}) and residual from the observed profiles (\textit{right}).
The \textit{black symbols} indicate the observed Stokes profiles at the position indicated by the \textit{cross symbol} in Figure~\ref{F-flux_map}.
The \textit{red} and \textit{blue lines} represent the fitting results of the MILOS inversion with the default settings and with the ``global'' scattered-light profile and the magnetic filling factor equal to the MERLIN values, respectively.
The Stokes profiles are normalized by the reference continuum intensity same as in Figure~\ref{F-stray}.}
\label{F-profile}
\end{figure}

\begin{figure} 
\centerline{\includegraphics[width=6cm]{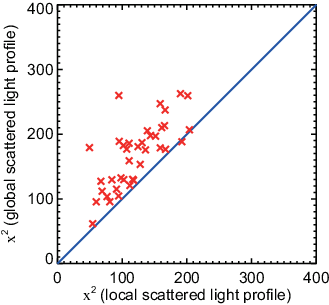}}
\caption{
Scatter plots for the $\chi^2$ values of the MILOS inversion with the ``local'' and ``global'' scattered-light profiles.
The ``local'' and ``global'' scattered-light profiles are used in the MILOS and MERLIN inversions with their default setting, respectively.
In the case of ``global'' scattered-light profile, the magnetic filling factor is forced to be at the value of the MERLIN inversion. 
The \textit{solid blue line} denotes the instances where the two inversion results are equal.
The  $\chi^2$ values are plotted only for pixels having a large discrepancy in the magnetic filling factor between the MILOS and MERLIN inversions.
The criteria for selecting these pixels are described in the text.
}
\label{F-chisqr}
\end{figure}

\begin{table}
\caption{Example of fitting results of MILOS inversion.}
\label{T-fitting}  
\begin{tabular}{lcc}     
\hline                   
 & local $I_{\rm{scatt}}$ & global $I_{\rm{scatt}}$ \\
 & (default) &  \\
\hline
Field strength [gauss] & 1166.0 & 807.1 \\
Inclination angle [degree]  & 62.2 & 70.0 \\
Azimuth angle [degree] & 116.8 & 118.1 \\
Filling factor & 0.34 & 0.87 \\
Doppler shift [km s$^{-1}$] &0.31 &0.31 \\
Doppler width &0.041 &0.044 \\
Damping parameter &0.14 &0.20 \\
Ratio of line center to continuum opacity &12.0 &10.9 \\
Surface value of the source function &0.44 &0.41 \\ 
Slope of the source function &1.14 &0.93 \\ 
\hline
$\chi^2$ & 201 & 282 \\
\hline
\end{tabular}
\end{table}

The observed Stokes profiles are compared with the fitting results of the MILOS inversion with the ``local'' and ``global'' scattered-light profiles (Figure~\ref{F-profile}).
This example corresponds to the strong magnetic patch indicated by the red symbol in Figure~\ref{F-flux_map}.
As mentioned in Section~\ref{S-ff}, the ``local'' scattered-light profiles are used for the MILOS inversion with the default setting.
In the case of the ``global'' scattered-light profile, its shift in the wavelength direction and the magnetic filling factor are forced to be at the values derived with the MERLIN inversion for each pixel.
This means that the difference in the fitting results is due to the difference in the assumptions of the scattered-light profile (see Section~\ref{S-ff}). 
The right panels in Figure~\ref{F-profile} show that the residual from the observed Stokes profiles is smaller with the MILOS default setting than that with the ``global'' scattered-light profiles for linear polarization (Stokes Q and U), but the residual is comparable between the two cases for circular polarization (Stokes V).
The difference in the results with the two assumptions of the scattered-light profiles (see Section~\ref{S-ff}) is mainly due to the fitting to the Stokes QU profiles rather than to Stokes V.
In Table~\ref{T-fitting},  the inversion result for this example with the MILOS default setting (left column) is compared to that with the ``global'' scattered-light profile (right column).
The $\chi^2$ value is estimated by the following equation:
\begin{eqnarray}   
    \chi^2 &=& \frac{1}{N} \sum_{i} \sum_{j}\frac{w_i(I_i^{\rm{obs}}(j)-I_i^{\rm{fit}}(j))^2}{\sigma^2},
                       \label{Eq-chisqr}     
\end{eqnarray}
where $N$ is the number of degrees of freedom, $i$ represents the Stokes IQUV parameters, $j$ represents the wavelength positions, $w_i$ is the weights of the Stokes profiles (Section~\ref{S-weights}), $I^{\rm{obs}}$ is the observed Stokes profiles, and $I^{\rm{fit}}$ is the fitting result.
The error ($\sigma$) is assumed to be 0.001 for all pixels.
The difference in the degrees of freedom ($N$) between the two settings is considered for the $\chi^2$ estimation.
The smaller residual from the observed profiles in the  MILOS inversion with the default setting is confirmed by the better $\chi^2$ value.

Table~\ref{T-fitting} shows that a more vertical magnetic field with the stronger field strength is obtained with the ``local'' scattered-profile, and this is consistent with the trends in Figure~\ref{F-vector}.
The magnetic filling factor, which is closely related to the assumption of a scattered-light profile considering its definition, is 0.87 for the ``global" scattered-light profile and 0.34 for the ``local'' scattered-light profile at this position. A possible explanation for the larger magnetic filling factor with the ``global'' scattered-light profile is caused by the intensity gap between the observed profile and the scattered-light profile. The continuum intensity of the observed Stokes I profile (top right plot in Figure~\ref{F-profile}) is similar to the ``local'' scattered-light profile in the MILOS inversion, but it is approximately 20 \%  brighter than the continuum intensity of the ``global'' scattered-light profile in the MERLIN inversion (Figure~\ref{F-stray}). As the magnetic filling factor becomes smaller in the MERLIN inversion, the continuum intensity of the fitting result deviates from the observed value. The continuum intensity and the line depth are determined not only by the contamination of the scattered-light profile but also by the other fitting parameters. For example, the slope and surface value of the source function and the ratio of the line center to the continuum opacity are smaller in the case of the ``global'' scattered-light profile. These smaller values tend to compensate for the gap in the intensities between the observed profile and the scattered-light profile.

We compare the $\chi^2$ values for the pixels with a large discrepancy in the magnetic filling factor, as in the example shown in Table~\ref{T-fitting}.
The magnetic filling factor is closely related to the assumptions of the scattered-light profiles (Equation~\ref{E-ff}) and its difference in the two assumptions is larger than the other parameters at least in this example.
Our selection criterion is a pixel having a filling factor greater than 0.8 in the MERLIN inversion and less than 0.4 in the MILOS inversion with the default setting.
We select only the areas of strong polarization ($>0.01$) to avoid the effects of noise as much as possible. 
As a result, 39 pixels are selected. 
Figure~\ref{F-chisqr} shows that the MILOS inversion results with the default setting have better or equal $\chi^2$ values for all 39 pixels.
This suggests that the assumption of the ``local'' scattered-light profiles provides better fitting results of the Stokes profiles.

\subsection{Dependence of Weights on Stokes Parameters}
\label{S-weights}

\begin{table}
\caption{Weights for Stokes IQUV in the inversion.}
\label{T-weight}  
\begin{tabular}{cccccl}     
\hline                   
Weight \# & I & Q & U & V & Note\\
\hline
0 & 1 & 20  & 20 & 10 & default values for MILOS \\
1 & 1 & 10  & 10 & 10 & \\
2 & 1 & 100 & 100 & 10 & default values for MERLIN \\
\hline
\end{tabular}
\end{table}


\begin{figure} 
\centerline{\includegraphics[width=12cm]{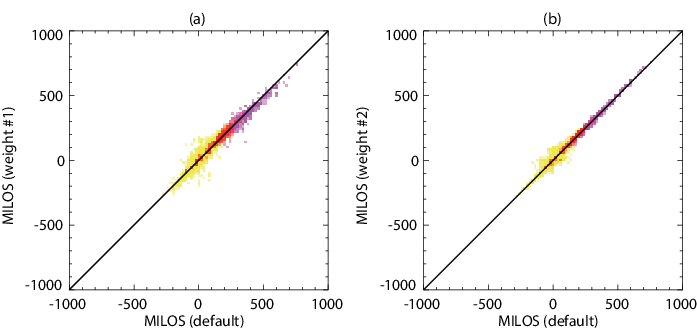}}
\caption{Same as Figure~\ref{F-flux}, but for the comparison between the MILOS inversion results with different weight values for the Stokes profiles.
The weight values are summarized in Table~\ref{T-weight}.}
\label{F-weight}
\end{figure}

The default weights for the Stokes IQUV profiles are different between the MILOS and MERLIN inversions, as shown in Table~\ref{T-weight}.
This difference may be due to the target of the database: the MERLIN inversion as used for the HAO level-2 database provides the data set on the whole Sun, and default MILOS inversion code as used for the production of the ISEE database is optimized for the polar observations.
The typical ratios of linear to circular polarization near the disk center and near the poles are different.
Figure~\ref{F-weight} shows that the radial magnetic-flux density does not change significantly in the areas of strong polarization even if we change the weights in the MILOS inversion from the default value to the MERLIN value.
Furthermore, similar results are obtained by the MILOS inversion with smaller weights for the linear polarization (weight \#2 in Table~\ref{T-weight}).
As mentioned in Section~\ref{S-ff}, the larger magnetic-flux density in the MERLIN inversion is related to the different assumptions of the scattered-light profiles.
The weight for the Stokes I profile is much smaller than those of the other Stokes parameters in all the cases listed in Table~\ref{T-weight}.
The effect caused by the mismatch between the observed Stokes I profile and the scattered-light profile can be reduced by the small weight to the fitting of the Stokes I profile.
Although the ratio of linear to circular polarization in the weights is important for obtaining weak magnetic fields comparable to noise, it is not a major cause for producing the difference in the magnetic-flux density in the areas of strong polarization.

\section{Summary and Discussions}
\label{S-Summary}
We find that the MERLIN inversion results tend to have approximately 7\%-10\% larger radial magnetic-flux density than the MILOS inversion results in the polar regions observed with Hinode/SOT-SP.
This trend is noticed in the polar magnetic fields observed at different times of the solar cycle.
However, the up to 10\% difference in magnetic-flux density does not affect the main conclusions regarding the key findings of the polar magnetic fields by Hinode, as mentioned in Section~\ref{S-Intro}.
These results imply that both inversions provide us consistent results for the polar magnetic flux over one solar cycle, although their results have a systematic difference of up to 10\%.

The radial magnetic-flux density derived from the two inversions becomes almost equal when the MILOS inversion is performed with the ``global'' scattered-light profiles and the magnetic filling factor same as the MERLIN value.
This suggests that the discrepancy in the magnetic-flux density is mainly caused by the different assumptions of the scattered-light profiles by the two inversions with their default settings (see Section~\ref{S-ff}).
The different assumptions of the scattered-light profiles also cause the discrepancy in the magnetic filling factor and the magnetic filling factor obtained with the MERLIN inversion tends to be larger.
One possible explanation of this trend is owing to a gap in the continuum intensity between the scattered-light profile and the observed Stokes I profile.
The continuum intensity of the observed Stokes I varies with heliocentric angles in the polar region because of limb darkening.
The intensity level of the ``global'' scattered-light profile, which is averaged over the field-of-view in the MERLIN inversion, differs from that of the observed Stokes I profile, except for particular areas.
This intensity gap induces fitting errors, so that the magnetic filling factor is limited to larger values (i.e. with smaller stray light contributions).
However, the intensity gap in the MILOS inversion is small because it uses the ``local'' scattered-light profile, which is averaged over the neighboring pixels.
The above discussion on discrepancy of the magnetic filling factor may be over-simplified because additional fitting parameters might be able to compensate for the continuum deviations. Nevertheless,  the assumption of the ``local'' scattered-light profile provides better fitting results at least in the pixels with the the large discrepancy of the magnetic filling factor. 
This is confirmed by the smaller $\chi^2$ values in the MILOS inversion with the default setting.
However, the smaller $\chi^2$ values do not simply imply that physical quantities can be always obtained more accurately. 
The possible problem of the inversion with the ``local'' scattered-light profiles was discussed in \cite{2011ApJ...731..125A}. From this point of view, it is important to investigate the deviation of the results of the two inversions from the ground truth in the polar region \citep{2023ApJ...951...23C}.

The radial component of the magnetic-field vector ($B_r$) is dominated by the line-of-sight magnetic field near the disk center and the contribution to $B_r$ from the transverse magnetic field increases closer to the limb.
In the following  we discuss the relationship between the magnetic filling factor and the Stokes profiles in the weak-field approximation to simplify the discussion, although the Milne-Eddington inversion codes are compared in this study.
The amplitude of the Stokes V profile has a linear relationship ($fB_l$) with the magnetic filling factor ($f$) and line-of-sight magnetic field ($B_l$) in the weak-field approximation \citep{2004ASSL..307.....L}.
The estimation of the magnetic-flux density at the disk center is comparatively less sensitive to the difference in the magnetic filling factor because this difference is compensated by the line-of-sight magnetic field to maintain the fit to the Stokes V profile.
The amplitudes of the Stokes Q and U profiles have a linear relationship with the magnetic filling factor ($f$), but a nonlinear relationship with the transverse magnetic field ($B_t$) in the weak-field approximation ($f\sqrt{B_t}$).
Therefore, the difference in the magnetic filling factor between the two inversions affects the radial magnetic-flux density in the polar observations, such as HOP 206.
Although the Stokes V residuals are similar in the fittings of the two inversions, the Stokes Q and U residuals are different between the two inversions.
The magnetic filling factor is often assumed to be 1 for the estimation of the magnetic-flux density \citep[e.g., VFISV for HMI,][]{2011SoPh..273..267B}.
The inclusion of the magnetic filling factor in the inversion code \citep{2021ApJ...923...84G} helps to improve the fitting to Stokes profiles and to mitigate the bias caused by a mismatch between of the line-of-sight and transverse components of the vector magnetic fields \citep{2022SoPh..297...17L, 2022SoPh..297..121L}. 
\citet{2021JSWSC..11...14P} provides the observational evidence that the estimation of the magnetic filling factor in combination with the azimuth ambiguity affects the orientation of the vector magnetic field, especially in the weak magnetic field regions.
Our results suggest that a better estimation of the magnetic filling factor is also important for the magnetic-flux density in the polar region.
For a better estimation of the magnetic filling factor, there is a possibility to improve both of the assumptions of the scattered-light profiles in the polar regions. For the ``global'' scattered-light profile, the gap in the continuum intensity between the observed Stokes I profile and the scattered-light profile can be reduced if the scattered-light profile is modified by a scaling factor to account for the limb darkening. For the ``local'' scattered-light profile, the averaging area of the 7x7 pixels may be small for polar faculae. In this case, the most of the 7x7 pixels are covered by the magnetic elements and the scattered-light profiles include the magnetic component.

The high-precision spectropolarimetric measurements with Hinode/SOT-SP significantly improve the estimation of the polar magnetic fields, but it remains an important frontier in solar physics.
\cite{2022ApJ...941..142P} suggested that the radial component of the magnetic field is underestimated by using the transverse field in the polar region, because the sensitivity of the transverse magnetic field is lower than that of the line-of-sight magnetic field.
Our results suggest that the estimation of the magnetic filling factor is also affected by the distance from the limb and it contributes to the radial magnetic-flux density.
However, this effect is not simply related to the distance from the limb.
For example, in the case of assuming the ``global'' scattered-light profile, the deviation of the continuum intensity from the averaged scattered-light profile becomes larger both at smaller heliocentric angles (i.e., the bottom edge of Figure~\ref{F-vector_map}) and at larger heliocentric angles close to the limb.
To compensate for the overestimation of the magnetic filling factor,  the transverse magnetic field is reduced more than the line-of-sight magnetic field.
This results in larger $B_r$ at smaller heliocentric angles ($B_r{\sim}B_l$) and smaller $B_r$ at larger heliocentric angles close to the limb ($B_r{\sim}B_t$).
In addition, the effects related to limb darkening (i.e., contamination of the scattered-light and lower field strength due to the observations in the upper layer) most likely complicate the estimation of the magnetic-flux density.
The observations of the polar magnetic fields from outside the ecliptic plane, such as the Polarimetric and Helioseismic Imager on the Solar Orbiter mission \citep[SO/PHI:][]{2020A&A...642A..11S} are important to derive the polar magnetic fields without these effects related to the distance from the limb.
The high spatial resolution and high precision polar magnetic-field measurements of Hinode/SOT-SP over one solar cycle are currently unique and invaluable data sets.
Another possible way to assess the uncertainties in the estimation of the polar magnetic flux would be to compare the Hinode/SOT-SP data with other spectropolarimetric data: long-term full disk observations \citep[e.g.,][]{2022ApJ...941..142P}, observations with near-infrared lines more sensitive to the weaker magnetic field \citep{2018A&A...616A..46P}, or observations with higher spatial resolution \citep{2020A&A...635A.210P, 2020A&A...644A..86P}.
In the near future, more superior data sets will be obtained with DKIST and other advanced instruments, but acquiring the data over one solar cycle is time consuming. 
Therefore, it is necessary to make effective use of the data obtained with Hinode/SOT-SP in order to advance our understanding of the polar magnetic field.

%

%

%

%
\begin{acks}
We acknowledge the anonymous reviewer for helpful suggestions and comments.
Hinode is a Japanese mission developed and launched by ISAS/JAXA, collaborating with NAOJ as a domestic partner, NASA and STFC (UK) as international partners. Scientific operation of the Hinode mission is conducted by the Hinode science team organized at ISAS/JAXA. This team mainly consists of scientists from institutes in the partner countries. Support for the post-launch operation is provided by JAXA and NAOJ(Japan), STFC (U.K.), NASA, ESA, and NSC (Norway).
\end{acks}

\begin{fundinginformation}
This work was carried out by the joint research program of the Institute for Space–Earth Environmental Research (ISEE), Nagoya University.
M.K. and Y.K. acknowledges JSPS KAKENHI Grant Numbers JP20KK0072, JP18H05234, and JP23K25916.
D.S. acknowledges JSPS KAKENHI Grant Numbers JP19K23472.
DOS acknowledges financial support from the grants CEX2021-001131S, PID2021-125325OB-C51, and PCI2022-135009-2, funded by MCIN/AEI/ 10.13039/501100011033, and by “ERDF A way of making Europe”, by the “European Union''.
This material is based upon work supported by the National Center for Atmospheric Research, which is a major facility sponsored by the National Science Foundation under Cooperative Agreement No. 1852977. N.N. and R.C. acknowledges support from NASA LWS Award 80NSSC20K0217.
\end{fundinginformation}

%
%
%
%
%
%
%

%
%
\bibliographystyle{spr-mp-sola}
\bibliography{sola_bibliography}  
%
%
%
%

\end{document}